\documentclass[3p,times,twocolumn]{elsarticle}

\usepackage{ecrc}

\def\lsno{La$_{2-x}$Sr$_x$NiO$_4$}
\def\lsco{La$_{2-x}$Sr$_x$CuO$_4$}
\def\lbco{La$_{2-x}$Ba$_x$CuO$_4$}

\def\lnsco{La$_{1.6-x}$Nd$_{0.4}$Sr$_x$CuO$_4$}
\def\lesco{La$_{1.8-x}$Eu$_{0.2}$Sr$_x$CuO$_4$}
\def\ybco{YBa$_2$Cu$_3$O$_{6+x}$}

\def\bscco{Bi$_2$Sr$_2$CaCu$_2$O$_{8+\delta}$}

\volume{00}

\firstpage{1}

\journalname{Physica B: Condensed Matter}

\runauth{J.M. Tranquada}

\jid{physb}

\jnltitlelogo{Physica B}

\usepackage{amssymb}

\usepackage[figuresright]{rotating}

\begin{document}

\begin{frontmatter}



\dochead{}

\title{Exploring intertwined orders in cuprate superconductors}


\author{John M. Tranquada\corref{cor1}}

\address{Condensed Matter Physics \&\ Materials Science Department\\
Brookhaven National Laboratory, Upton, NY 11973-5000, USA}

\cortext[cor1]{E-mail address: jtran@bnl.gov}

\begin{abstract}
The concept of intertwined orders has been introduced to describe the cooperative relationship between antiferromagnetic spin correlations and electron (or hole) pair correlations that develop in copper-oxide superconductors.  This contrasts with systems in which, for example, charge-density-wave (CDW) order competes for Fermi surface area with superconductivity.  \lbco\ with $x=0.125$ provides an example in which the ordering of spin stripes coincides with the onset of two-dimensional superconducting correlations.  The apparent frustration of the interlayer Josephson coupling has motivated the concept of the pair-density-wave superconductor, a state that theoretical calculations show to be energetically competitive with the uniform $d$-wave superconductor.  Even at $x=0.095$, where there is robust superconductivity below 32~K in zero field, the coexistence of strong, low-energy, incommensurate spin excitations implies a spatially modulated and intertwined pair wave function.  Recent observations of CDW order in \ybco\ and other cuprate families have raised interesting questions regarding the general role of charge modulations and the relation to superconductivity.  While there are differences in the doping dependence of the modulation wave vectors in \ybco\ and \lbco, the maximum ordering strength is peaked at the hole concentration of 1/8 in both cases.  There are also possible connections with the quantum oscillations that have been detected about the same hole concentration but at high magnetic fields.  Resolving these relationships remains a research challenge.
\end{abstract}


\end{frontmatter}


\section{Competing vs.\ intertwined orders}
\label{}

Underdoped cuprate superconductors exhibit a ``pseudogap'' phase that becomes apparent below a temperature $T^*$, where $T^*$ decreases with increased doping, approaching the superconducting transition temperature $T_c$ for hole concentrations beyond optimal (where optimal doping corresponds to maximum $T_c$) \cite{norm05}.  It has become common to discuss ``competing'' orders that may be responsible for the pseudogap and which may compete with superconductivity.   There are certainly many good examples of systems where the transition temperature for one type of order, such as charge density wave (CDW) or spin density wave (SDW) order, drops to zero at the same point where the superconducting $T_c$ reaches its maximum.  For example, pressure causes the CDW transitions in TaS$_2$ to decrease from 550~K towards zero, while the superconducting $T_c$ rises to 5~K, and continues in the absence of CDW order at higher pressures \cite{sipo08}.  At this meeting, N\'u\~nez-Regueiro reviewed a variety of related systems in which an order such as CDW or SDW competes for Fermi surface area with superconductivity \cite{nune14}.

A rather different sort of picture has emerged from studies of \lbco\ \cite{li07,tran08,huck11}.  Here, SDW and CDW orders appear to be intertwined with a spatially-modulated superconductivity, referred to as a pair-density-wave (PDW) superconductor \cite{hime02,berg07,berg09b}.   Antiferromagnetic and pairing correlations seem to develop in a cooperative fashion, forming a spatially self-organized pattern.  Such behavior has motivated the perspective that intertwined orders are a common feature of the cuprate superconductors \cite{frad14}.  

Recent observations of CDW order in \ybco\ \cite{wu11,ghir12,chan12a} have led to renewed discussions of competing order.   While many questions remain, it appears that the concept of intertwined order is relevant to the interpretation of these results \cite{frad14}.

\section{Pseudogap}

The nominal Fermi surface of an underdoped CuO$_2$ plane forms a hole pocket about ${\bf k}=(\pi/a,\pi/a)$.  In the superconducting state, the $d$-wave pairing gap has a node near $(\pi/2a,\pi/2a)$, with the gap maxima occurring in the vicinity of $(\pi/a,0)$ and $(0,\pi/a)$ (known as the antinodal regions).   In the temperature range above $T_c$ but below $T^*$, the superconducting gap closes along a finite arc about the nodal point, but a gap remains in the antinodal region \cite{ding96,yang11,vish12}.  The energy scale of the pseudogap is quite similar to that of the superconducting gap, just as the magnitude of $T^*$ is similar to $T_c$.  

Is the pseudogap associated with some type of competing order?  There have certainly been proposals of various specific orders intended to explain the pseudogap \cite{varm06,chak01}.   In a conventional picture, a well-defined Fermi liquid would be present at $T>T^*$, and the pseudogap would appear and grow as the competing order develops.  A problem with such a scenario is that there is no Fermi liquid state at $T>T^*$.  Optical conductivity studies demonstrate the absence of a Drude peak, and hence the absence of coherent quasiparticles \cite{lee05,take03}.  The appearance of coherent states, associated with the nodal arc states, occurs only on cooling below $T^*$.  This happens in parallel with the development of antiferromagnetic spin correlations, as indicated by the bulk spin susceptibility \cite{huck08,gork06}.  Thus, charge coherence and spin correlations appear to develop at the same time, suggesting cooperation rather that competition.

\section{\lbco\ and the PDW}

The case of \lbco\ provides an example of how spins and charges can organize themselves in an intertwined fashion that enables strong pairing of holes.  After the initial discovery of high-temperature superconductivity \cite{bedn86}, exploration of the phase diagram revealed a sharp minimum of $T_c$ at the hole concentration $x\approx 1/8$ \cite{mood88}.  Such a strong dip in $T_c$ is not observed in chemically-similar \lsco\ \cite{naga93}.  It was quickly demonstrated that the unusual behavior in \lbco\ is associated with a phase transition to the low-temperature tetragonal (LTT) phase at a temperature of $\sim60$~K \cite{axe89}.  When single crystals became available, neutron diffraction measurements on the related system La$_{1.48}$Nd$_{0.4}$Sr$_{0.12}$CuO$_4$, which also exhibits the LTT phase, found the presence of charge and spin stripe order \cite{tran95a,ichi00}.  The occurrence of stripe order was eventually confirmed in \lbco\ \cite{fuji04,huck11} and also in \lesco\ \cite{fink11}.  Despite the absence of bulk LTT order, very recent x-ray scattering measurements have demonstrated the presence of charge stripe order in La$_{1.88}$Sr$_{0.12}$CuO$_4$ \cite{wu12,chri14,crof14,tham14}, where spin stripe order had previously been detected \cite{kimu00}.

The occurrence of charge and spin stripe order was predicted long before it was detected  \cite{zaan89,mach89,poil89,schu90}.  These analyses properly captured the fact that the charge stripes act as antiphase domain walls for the spin stripes, so that wave vectors ${\bf q}_c$ for charge order and ${\bf q}_s$ for spin order (with the latter measured relative to ${\bf Q}_{\rm AF}=(\pi/a,\pi/a)$) are related by ${\bf q}_c=2{\bf q}_s$.  Where they disagree with experiment is in predicting that the charge stripes should be insulating, with one hole per Cu site along a stripe, and with a period $d_c = 2\pi/(q_ca) = 1/x$; experiment finds $d_c\approx 2/x$ \cite{tran97a,huck11}, corresponding to about half of a hole per Cu site along a stripe.  After stripes were detected, the role of stripe fluctuations was considered in order to account for metallic behavior in cuprates \cite{zaan96a}.   [The existence of charge-stripe fluctuations in the analog system \lsno\ (with insulating charge stripes) has now been confirmed by neutron scattering experiments \cite{abey13,anis14} (and the system becomes a bad metal when the stripes are fluctuating \cite{kats96,home03}).]

An alternative theoretical approach is the concept of Coulomb-frustrated phase separation \cite{emer93,kive94,low94}.  An early analysis of the $t$--$J$ model indicated that a low density of holes in an antiferromagnet would tend to phase separate \cite{emer90}; the phase separation is possible because the model does not take account of extended Coulomb interactions.  Accounting for such interactions with an effective Hamiltonian led to predictions of a variety of charge-ordered states, including stripe and checkerboard orders \cite{low94}.  After stripes were experimentally observed, the possibility of superconductivity arising from pairing correlations within the hole-rich stripes was explored \cite{emer97}.   At the time, only in-phase coupling of pairing between stripes was considered, and it was concluded that superconducting order would require fluctuating stripes in order to avoid competition from CDW order \cite{kive98}.

The formation of stripes can be viewed as a competition between the kinetic energy of the holes and the exchange energy between the spins.  This effect has been efficiently captured by density matrix renormalization group (DMRG) calculations of the $t$--$J$ model on finite systems \cite{whit98a,whit98c,scal12b}.  These calculations yield a charge density of $\sim0.5$ hole/site, consistent with experiment.  Furthermore, strong pairing correlations within the stripes are obtained \cite{whit09}.

Experimentally, stripe order in \lbco\ with $x=1/8$ strongly depresses bulk superconducting order; however, careful studies of the anisotropic resistivity and magnetic susceptibility in single crystals revealed evidence for the onset of two-dimensional (2D) superconducting correlations at $\sim40$~K, simultaneously with spin stripe order \cite{li07,tran08}.  (Similar observations of 2D superconductivity have been reported for stripe-ordered \lnsco\ \cite{ding08}.)   While superconductivity in the cuprates is always driven by the electronic correlations within the CuO$_2$ planes, there is inevitably a Josephson coupling between the layers that induces 3D superconducting order when the correlation length for 2D superconducting phase order becomes sufficiently large.  The suppression of 3D order in \lbco\ $x=1/8$ suggests that the interlayer Josephson coupling must be frustrated.  This frustration effect was first observed in a $c$-axis optical conductivity study of La$_{1.85-y}$Nd$_y$Sr$_{0.15}$CuO$_4$, where the increasing Nd concentration leads to a transition to the LTT phase and stripe order, together with a collapse of the Josephson plasma resonance, despite the survival of bulk superconductivity \cite{taji01}.  These observations motivated the concept of the PDW \cite{hime02,berg07,berg09b}.  Because of the rotation of the charge-stripe direction by 90$^\circ$ between layers in the LTT phase \cite{vonz98}, Josephson coupling between the PDW order in neighboring layers is zero.  It is interesting to note that the PDW state has also been proposed from a different starting point, that of ``Amperean pairing'' \cite{lee14}.

While the PDW state provides a good explanation of the 1/8 anomaly, is it an energetically competitive state?  Recent variational tensor network studies of the $t$--$J$ model suggest that it is \cite{corb11,corb14}.  Using reasonable parameters, the stripe state with antiphase $d$-wave superconducting order is very close to that of stripes with in-phase $d$-wave order; both are found to be slightly lower in energy than the uniform $d$-wave state.  There has also been an interesting mean-field analysis of the PDW state \cite{lode11}.

Is the PDW state relevant at hole concentrations away from $x=1/8$?  Optical conductivity studies of \lsco\ with $x=0.1$ provided evidence that superconducting layers become decoupled in a $c$-axis magnetic field \cite{scha10,scha10b}.  This led us to investigate the behavior of \lbco\ with $x=0.095$, for which the zero-field $T_c$ is 32~K.  Measurements of resistivity parallel and perpendicular to the planes showed that the in-plane resistivity remains effectively zero while the $c$-axis resistivity can become finite in quite modest $c$-axis fields ($<1$~T) \cite{wen12}.  Such measurements have been extended to fields as high as 35~T, at which point the in-plane resistivity is negligible at 13~K while the $c$-axis resistivity is compatible with strongly insulating behavior \cite{steg13}.  While the stability of this novel high-field state is yet to understood, these results suggest that the PDW state may be induced by a $c$-axis magnetic field, leading to the frustration of the interlayer Josephson coupling.

Can the modulated superconducting state be relevant to good superconductors at zero field?  We have recently investigated the spin fluctuations in the \lbco\ $x=0.095$ sample by inelastic neutron scattering \cite{xu14}.  This sample exhibits weak spin and charge stripe order in the superconducting state.  Measurements of the incommensurate spin fluctuations at $T\ll T_c$ reveal gapless excitations with a strength comparable to that of spin waves in antiferromagnetic La$_2$CuO$_4$.  Similarly, there is no sign of a ``spin resonance'' peak.  The absence of a spin gap and resonance peak violates the current paradigm for antiferromagnetic superconductors \cite{scal12a,esch06}. 

Empirically, superconductivity does not like to coexist with static or quasi-static local antiferromagnetic order.  Near optimum doping, the spectral weight of low-energy spin fluctuations is weak in the normal state, and below $T_c$ the system may organize itself so as to gap the spin fluctuations below the energy of the superconducting gap.  In the case of \lbco\ with $x=0.095$, we have bulk superconductivity coexisting with spatially-modulated quasi-static antiferromagnetic spin correlations.  These observations seem to imply that the superconducting pair wave function must be spatially modulated, to minimize overlap with the low-energy spin correlations \cite{xu14}.  The modulated state in zero field is probably not a pure PDW, since the PDW is quite sensitive to disorder and would tend to frustrate the interlayer Josephson coupling; nevertheless, some degree of modulation appears unavoidable.

\section{CDW in \ybco}

Soon after quantum oscillations were observed in underdoped \ybco\ at high magnetic fields and low temperature \cite{lebo07,doir07,seba08,rigg11}, it was proposed that the corresponding Fermi-surface pockets might result from a reconstruction of the Fermi surface due to spin stripes \cite{mill07} or charge stripes \cite{yao11}.  Similarities were demonstrated in the temperature dependence of transport properties, such as the thermoelectric power, between \ybco\ (at finite magnetic field) and stripe-ordered ``214'' cuprates \cite{chan10}.   Nuclear magnetic resonance (NMR) studies revealed a magnetic-field-induced splitting of NMR lines, with an onset at a minimum field of 10~T, consistent with some type of charge order that appears at high field \cite{wu11,wu13a}.  A sound-velocity study also provides evidence for a transition to a distinct high-field state at low temperature (although the transition field seems to be different from that found by NMR) \cite{lebo13}.  Thermal conductivity measurements provide further support for such a transition, and torque magnetometry provides evidence that diamagnetism is present above that transition (although this is above the vortex melting transition) \cite{yu14}.  It has been proposed that the transition between states with finite diamagnetism may correspond to a switch from uniform $d$-wave to PDW superconductivity \cite{yu14}.

In the mean time, distinct short-range CDW order has been detected in \ybco\ by both resonant soft-x-ray scattering \cite{ghir12,achk12,blan13} and high-energy x-ray diffraction \cite{chan12a,blac13a}.  Results for a range of dopings have recently been summarized \cite{huck14,blan14}.   The most intense CDW scattering and the highest onset temperatures occur near a hole concentration of 0.12, the same point at which the upper critical field $H_{c2}$ (as measured by the onset of finite resistivity) shows a dip \cite{rams12,gris14}, and remarkably similar to the 1/8 anomaly in \lbco.   The maximum CDW onset temperature is $\sim150$~K, which is quite high compared to charge ordering in \lbco; however, to the extent that this ordering may depend on anisotropic Cu-O bonds, such an anisotropy is present in orthorhombic \ybco\ up to the Cu-O chain ordering temperature, above 300~K.  On cooling, the CDW intensity reaches a peak at the superconducting $T_c$, and falls below it \cite{ghir12,chan12a}; application of a $c$-axis magnetic field strong enough to depress the superconductivity causes an enhancement of the CDW intensity \cite{chan12a}.  (Similar behavior is seen in \lbco\ for $x$ far enough from 1/8 that the charge order is not close to saturation \cite{huck13}.)  

Diffraction studies have found similar short-range CDW correlations in Bi$_2$Sr$_{2-x}$La$_x$CuO$_{6+\delta}$ \cite{comi14}, \bscco\ \cite{dasi14}, and HgBa$_2$CuO$_{4+\delta}$ \cite{tabi14}.  (CDW order has even been seen in the electron-doped superconductor Nd$_{2-x}$Ce$_x$CuO$_4$ \cite{dasi14b}.) For the former two materials, it has been shown \cite{comi14,dasi14} that the CDW correlations correspond to modulations previously detected by scanning tunneling spectroscopy studies, especially at biases comparable to the pseudogap energy \cite{howa03b,kohs07,park10}.  A significant observation from many of these studies is that the CDW wave vector tends to decrease slowly with hole doping, in contrast to the relationship $q_c/a^* \approx 2x$ already noted for \lbco\ and closely related compounds.  The characteristic {\it spin} wave vectors for low-energy magnetic excitations in Bi$_{2+x}$Sr$_{2-x}$CuO$_{6+y}$ and \ybco\ {\it increase} with hole doping in a fashion quantitatively similar to \lbco\ and \lsco\ \cite{enok13}; it follows that the CDW modulations in \ybco\ and these other compounds has no direct connection to the spin correlations.  It is notable that that inducing static incommensurate spin order in \ybco\ through the partial substitution of Zn causes the CDW signal present without Zn to weaken \cite{blan13}; this result suggests that the high-temperature CDW correlations are distinct from the charge and spin stripes found in \lbco.

The connection between the CDW orders seen at low and high fields in \ybco\ has yet to be resolved.  NMR studies show that the modulation in the low-field CDW is relatively weak, as it only results in a small, temperature-dependent line broadening \cite{wu14}.  The same appears to be true in HgBa$_2$CuO$_{4+\delta}$, where, despite narrow $^{17}$O NMR lines, an early study found no sign of anything unusual \cite{bobr97}.   It seems likely that the high-field CDW is related to the modulations detected by scanning tunneling spectroscopy within halos about magnetic vortices in \bscco\ \cite{hoff02}; these modulations have been detected via resonances at an energy smaller than the superconducting gap.  Further studies of this kind may provide a better understanding.

For \ybco, a plot of $T_c$ measured in a $c$-axis field of 15 or 30~T shows a doping dependence with a strong dip at the hole concentration of 0.12 \cite{gris14}.  This two-hump behavior is quite similar to that seen in \lbco\ \cite{huck11}.  While the discussion so far has focused on the charge correlations that are optimized at 0.12, it is also interesting to consider the character of the robust superconducting state in the low-doping dome.  For \lbco, this corresponds to $x=0.095$, which was argued in the previous section to have a spatially-modulated superconducting wave function related to the PDW state.  For \ybco, the lower dome has a maximum at the hole concentration 0.08, where neutron scattering studies have found quasi-static, incommensurate spin correlations \cite{hink08}, with the elastic response enhanced by an applied magnetic field \cite{haug09}.  This behavior is quite similar to \lbco, and suggests the relevance of PDW-like superconductivity.

\section{Conclusion}

The intertwined orders of spin, charge, and superconductivity found in \lbco\ at $x=1/8$, where bulk superconductivity is strongly depressed, appear to be relevant also to understanding the robust superconductivity found at $x=0.095$.  While there may be differences with the type of CDW correlations found in other cuprate families, the fact that charge order is optimized near the hole concentration of 1/8 suggests a common ``ineluctable complexity'' \cite{frad12}.   Quasi-static incommensurate spin correlations are likely intertwined with superconductivity in \ybco\ with a hole concentration of 0.08, suggesting that such behavior may be common in the cuprates.  It will be of interest to test the generality of such behavior in future experiments, as well as to perform direct tests for PDW superconducting order \cite{berg09a}.



\section{Acknowledgments}

I am grateful to many experimental collaborators, to S. A. Kivelson and E. Fradkin for collaborations on interpretation of the experiments, and to A. V. Chubukov for useful discussions.
Work at Brookhaven is supported by the Office of Basic Energy Sciences, Division of Materials Science and Engineering, U.S. Department of Energy under Contract No.\ DE-AC02-98CH10886.  This paper was written in part while the author was in residence at the Kavli Institute for Theoretical Physics, which is supported in part by the National Science Foundation under Grant No.\ NSF PHY11-25915.




\begin{thebibliography}{10}
\expandafter\ifx\csname url\endcsname\relax
  \def\url#1{\texttt{#1}}\fi
\expandafter\ifx\csname urlprefix\endcsname\relax\def\urlprefix{URL }\fi
\expandafter\ifx\csname href\endcsname\relax
  \def\href#1#2{#2} \def\path#1{#1}\fi

\bibitem{norm05}
M.~R. Norman, D.~Pines, C.~Kallin, 
  Adv. Phys. 54 (2005) 715.

\bibitem{sipo08}
B.~Sipos, A.~F. Kusmartseva, A.~Akrap, H.~Berger, L.~Forro, E.~Tutis, 
Nat. Mater. 7 (2008) 960--965.

\bibitem{nune14}
M. N\'u\~nez-Regueiro, preprint (2014).

\bibitem{li07}
Q.~Li, M.~{H\"ucker}, G.~D. Gu, A.~M. Tsvelik, J.~M. Tranquada,
  Phys. Rev. Lett. 99 (2007) 067001.

\bibitem{tran08}
J.~M. Tranquada, G.~D. Gu, M.~H{\"u}cker, Q.~Jie, H.-J. Kang, R.~Klingeler,
  Q.~Li, N.~Tristan, J.~S. Wen, G.~Y. Xu, Z.~J. Xu, J.~Zhou, M.~v.~Zimmermann,
  Phys. Rev. B 78 (2008) 174529.

\bibitem{huck11}
M.~H\"ucker, M.~v.~Zimmermann, G.~D. Gu, Z.~J. Xu, J.~S. Wen, G.~Xu, H.~J.
  Kang, A.~Zheludev, J.~M. Tranquada, 
  Phys. Rev. B 83 (2011) 104506.

\bibitem{hime02}
A.~Himeda, T.~Kato, M.~Ogata, 
  Phys. Rev. Lett. 88 (2002) 117001.

\bibitem{berg07}
E.~Berg, E.~Fradkin, E.-A. Kim, S.~A. Kivelson, V.~Oganesyan, J.~M. Tranquada,
  S.~C. Zhang, 
  Phys. Rev. Lett. 99 (2007) 127003.

\bibitem{berg09b}
E.~Berg, E.~Fradkin, S.~A. Kivelson, J.~M. Tranquada, 
  New J. Phys. 11 (2009) 115004.

\bibitem{frad14}
E.~Fradkin, S.~A. Kivelson, J.~M. Tranquada, {Theory of Intertwined Orders in
  High Temperature Superconductors}, arXiv:1407.4480 (2014).

\bibitem{wu11}
T.~Wu, H.~Mayaffre, S.~Kramer, M.~Horvatic, C.~Berthier, W.~N. Hardy, R.~Liang,
  D.~A. Bonn, M.-H. Julien, 
  Nature 477 (2011) 191--194.

\bibitem{ghir12}
G.~Ghiringhelli, M.~Le~Tacon, M.~Minola, S.~Blanco-Canosa, C.~Mazzoli, N.~B.
  Brookes, G.~M. De~Luca, A.~Frano, D.~G. Hawthorn, F.~He, T.~Loew, M.~M. Sala,
  D.~C. Peets, M.~Salluzzo, E.~Schierle, R.~Sutarto, G.~A. Sawatzky,
  E.~Weschke, B.~Keimer, L.~Braicovich, 
  Science 337 (2012) 821--825.

\bibitem{chan12a}
J.~Chang, E.~Blackburn, A.~T. Holmes, N.~B. Christensen, J.~Larsen, J.~Mesot,
  R.~Liang, D.~A. Bonn, W.~N. Hardy, A.~Watenphul, M.~v. Zimmermann, E.~M.
  Forgan, S.~M. Hayden, 
  Nat. Phys. 8 (2012) 871--876.

\bibitem{ding96}
H.~Ding, T.~Yokoya, J.~C. Campuzano, T.~Takahashi, M.~Randeria, M.~R. Norman,
  T.~Mochiku, K.~Kadowaki, J.~Giapintzakis, Nature 382 (1996) 51.

\bibitem{yang11}
H.-B. Yang, J.~D. Rameau, Z.-H. Pan, G.~D. Gu, P.~D. Johnson, H.~Claus, D.~G.
  Hinks, T.~E. Kidd, 
  Phys. Rev. Lett. 107 (2011) 047003.

\bibitem{vish12}
I.~M. Vishik, M.~Hashimoto, R.-H. He, W.-S. Lee, F.~Schmitt, D.~Lu, R.~G.
  Moore, C.~Zhang, W.~Meevasana, T.~Sasagawa, S.~Uchida, K.~Fujita, S.~Ishida,
  M.~Ishikado, Y.~Yoshida, H.~Eisaki, Z.~Hussain, T.~P. Devereaux, Z.-X. Shen,
  Proc. Natl. Acad. Sci. 109 (2012) 18332--18337.

\bibitem{varm06}
C.~M. Varma, 
Phys. Rev. B 73 (2006) 155113.

\bibitem{chak01}
S.~Chakravarty, R.~B. Laughlin, D.~K. Morr, C.~Nayak, Phys. Rev. B 63 (2001)
  094503.

\bibitem{lee05}
Y.~S. Lee, K.~Segawa, Z.~Q. Li, W.~J. Padilla, M.~Dumm, S.~V. Dordevic, C.~C.
  Homes, Y.~Ando, D.~N. Basov, Phys. Rev. B 72 (2005) 054529.

\bibitem{take03}
K.~Takenaka, J.~Nohara, R.~Shiozaki, S.~Sugai, Phys. Rev. B 68 (2003) 134501.

\bibitem{huck08}
M.~{H\"ucker}, G.~D. Gu, J.~M. Tranquada, 
  Phys. Rev. B 78 (2008) 214507.

\bibitem{gork06}
L.~P. Gor'kov, G.~B. Teitel'baum, 
  Phys. Rev. Lett. 97 (2006) 247003.

\bibitem{bedn86}
J.~Bednorz, K.~M\"uller, 
  Z. Phys. B 64 (1986) 189--193.

\bibitem{mood88}
A.~R. Moodenbaugh, Y.~Xu, M.~Suenaga, T.~J. Folkerts, R.~N. Shelton, Phys. Rev.
  B 38 (1988) 4596.

\bibitem{naga93}
T.~Nagano, Y.~Tomioka, Y.~Nakayama, K.~Kishio, K.~Kitazawa, 
  Phys. Rev. B 48 (1993) 9689--9696.

\bibitem{axe89}
J.~D. Axe, A.~H. Moudden, D.~Hohlwein, D.~E. Cox, K.~M. Mohanty, A.~R.
  Moodenbaugh, Y.~Xu, 
  Phys. Rev. Lett. 62 (1989) 2751.

\bibitem{tran95a}
J.~M. Tranquada, B.~J. Sternlieb, J.~D. Axe, Y.~Nakamura, S.~Uchida, 
  Nature 375 (1995) 561.

\bibitem{ichi00}
N.~Ichikawa, S.~Uchida, J.~M. Tranquada, T.~Niem\"oller, P.~M. Gehring, S.-H.
  Lee, J.~R. Schneider, 
  Phys. Rev. Lett. 85 (2000) 1738--1741.

\bibitem{fuji04}
M.~Fujita, H.~Goka, K.~Yamada, J.~M. Tranquada, L.~P. Regnault, 
  Phys. Rev. B 70 (2004) 104517.

\bibitem{fink11}
J.~Fink, V.~Soltwisch, J.~Geck, E.~Schierle, E.~Weschke, B.~B\"uchner, 
  Phys. Rev. B 83 (2011) 092503.

\bibitem{wu12}
H.~H. Wu, M.~Buchholz, C.~Trabant, C.~F. Chang, A.~C. Komarek, F.~Heigl, M.~v.
  Zimmermann, M.~Cwik, F.~Nakamura, M.~Braden, C.~Sch\"u{\ss}ler-Langeheine,
  Nat. Commun. 3 (2012) 1023.

\bibitem{chri14}
N.~B. Christensen, J.~Chang, J.~Larsen, M.~Fujita, M.~Oda, M.~Ido, N.~Momono,
  E.~M. Forgan, A.~T. Holmes, J.~Mesot, M.~H\"ucker, M.~v.~Zimmermann, {Bulk
  charge stripe order competing with superconductivity in
  La$_{2-x}$Sr$_x$CuO$_4$ ($x=0.12$)}, arXiv:1404.3192 (2014).

\bibitem{crof14}
T.~P. Croft, C.~Lester, M.~S. Senn, A.~Bombardi, S.~M. Hayden, 
  Phys. Rev. B 89 (2014) 224513.

\bibitem{tham14}
V.~Thampy, M.~P.~M. Dean, N.~B. Christensen, L.~Steinke, Z.~Islam, M.~Oda,
  M.~Ido, N.~Momono, S.~B. Wilkins, J.~P. Hill, 
  Phys. Rev. B 90 (2014) 100510.

\bibitem{kimu00}
H.~Kimura, K.~Hirota, C.-H. Lee, K.~Yamada, G.~Shirane, 
  J. Phys. Soc. Jpn. 69 (2000) 851--857.

\bibitem{zaan89}
J.~Zaanen, O.~Gunnarsson, 
  Phys. Rev. B 40 (1989) 7391.

\bibitem{mach89}
K.~Machida, 
Physica C 158 (1989) 192.

\bibitem{poil89}
D.~Poilblanc, T.~M. Rice, Phys. Rev. B 39 (1989) 9749.

\bibitem{schu90}
H.~J. Schulz, Phys. Rev. Lett. 64 (1990) 1445.

\bibitem{tran97a}
J.~M. Tranquada, J.~D. Axe, N.~Ichikawa, A.~R. Moodenbaugh, Y.~Nakamura,
  S.~Uchida, 
  Phys. Rev. Lett. 78 (1997) 338.

\bibitem{zaan96a}
J.~Zaanen, M.~L. Horbach, W.~van Saarloos, Phys. Rev. B 53 (1996) 8671.

\bibitem{abey13}
A.~M.~M. Abeykoon, E.~S. Bo\ifmmode~\check{z}\else \v{z}\fi{}in, W.-G. Yin,
  G.~Gu, J.~P. Hill, J.~M. Tranquada, S.~J.~L. Billinge, 
  Phys. Rev. Lett. 111 (2013) 096404.

\bibitem{anis14}
S.~Anissimova, D.~Parshall, G.~D. Gu, K.~Marty, M.~D. Lumsden, S.~Chi, J.~A.
  Fernandez-Baca, D.~L. Abernathy, D.~Lamago, J.~M. Tranquada, D.~Reznik,
  Nat. Commun. 5 (2014) 3467.

\bibitem{kats96}
T.~Katsufuji, T.~Tanabe, T.~Ishikawa, Y.~Fukuda, T.~Arima, Y.~Tokura, 
  Phys. Rev. B 54 (1996) R14230--R14233.

\bibitem{home03}
C.~C. Homes, J.~M. Tranquada, Q.~Li, A.~R. Moodenbaugh, D.~J. Buttrey,
  Phys. Rev. B 67 (2003) 184516.

\bibitem{emer93}
V.~J. Emery, S.~A. Kivelson, Physica C 209 (1993) 597.

\bibitem{kive94}
S.~A. Kivelson, V.~J. Emery, in: K.~S. Bedell, Z.~Wang, D.~E. Meltzer, A.~V.
  Balatsky, E.~Abrahams (Eds.), Strongly Correlated Electronic Materials: The
  Los Alamos Symposium 1993, Addison-Wesley, Reading, MA, 1994, pp. 619--656.

\bibitem{low94}
U.~L\"ow, V.~J. Emery, K.~Fabricius, S.~A. Kivelson, Phys. Rev. Lett. 72 (1994)
  1918.

\bibitem{emer90}
V.~J. Emery, S.~A. Kivelson, H.~Q. Lin, Phys. Rev. Lett. 64 (1990) 475.

\bibitem{emer97}
V.~J. Emery, S.~A. Kivelson, O.~Zachar, 
  Phys. Rev. B 56 (1997) 6120--6147.

\bibitem{kive98}
S.~A. Kivelson, E.~Fradkin, V.~J. Emery, 
  Nature 393 (1998) 550--553.

\bibitem{whit98a}
S.~R. White, D.~J. Scalapino, 
  Phys. Rev. Lett. 80 (1998) 1272--1275.

\bibitem{whit98c}
S.~R. White, D.~J. Scalapino, 
  Phys. Rev. Lett. 81 (1998) 3227--3230.

\bibitem{scal12b}
D.~Scalapino, S.~White, 
Physica C 481 (2012) 146--152.

\bibitem{whit09}
S.~R. White, D.~J. Scalapino, 
Phys. Rev. B 79 (2009) 220504.

\bibitem{ding08}
J.~F. Ding, X.~Q. Xiang, Y.~Q. Zhang, H.~Liu, X.~G. Li, 
  Phys. Rev. B 77 (2008) 214524.

\bibitem{taji01}
S.~Tajima, T.~Noda, H.~Eisaki, S.~Uchida, 
  Phys. Rev. Lett. 86 (2001) 500--503.

\bibitem{vonz98}
M.~v.~Zimmermann, A.~Vigliante, T.~Niem\"oller, N.~Ichikawa, T.~Frello,
  S.~Uchida, N.~H. Andersen, J.~Madsen, P.~Wochner, J.~M. Tranquada, D.~Gibbs,
  J.~R. Schneider, 
  Europhys. Lett. 41 (1998) 629.

\bibitem{lee14}
P.~A. Lee, 
  Phys. Rev. X 4 (2014) 031017.

\bibitem{corb11}
P.~Corboz, S.~R. White, G.~Vidal, M.~Troyer, 
  Phys. Rev. B 84 (2011) 041108.

\bibitem{corb14}
P.~Corboz, T.~M. Rice, M.~Troyer, 
  Phys. Rev. Lett. 113 (2014) 046402.

\bibitem{lode11}
F.~Loder, S.~Graser, A.~P. Kampf, T.~Kopp, 
  Phys. Rev. Lett. 107 (2011) 187001.

\bibitem{scha10}
A.~A. Schafgans, A.~D. LaForge, S.~V. Dordevic, M.~M. Qazilbash, W.~J. Padilla,
  K.~S. Burch, Z.~Q. Li, S.~Komiya, Y.~Ando, D.~N. Basov, 
  Phys. Rev. Lett. 104 (2010) 157002.

\bibitem{scha10b}
A.~A. Schafgans, C.~C. Homes, G.~D. Gu, S.~Komiya, Y.~Ando, D.~N. Basov,
  Phys. Rev. B 82 (2010) 100505.

\bibitem{wen12}
J.~Wen, Q.~Jie, Q.~Li, M.~H\"ucker, M.~v.~Zimmermann, S.~J. Han, Z.~Xu, D.~K.
  Singh, R.~M. Konik, L.~Zhang, G.~Gu, J.~M. Tranquada, 
  Phys. Rev. B 85 (2012) 134513.

\bibitem{steg13}
Z.~Stegen, S.~J. Han, J.~Wu, A.~K. Pramanik, M.~H\"ucker, G.~Gu, Q.~Li, J.~H.
  Park, G.~S. Boebinger, J.~M. Tranquada, 
  Phys. Rev. B 87 (2013) 064509.

\bibitem{xu14}
Z.~Xu, C.~Stock, S.~Chi, A.~I. Kolesnikov, G.~Xu, G.~Gu, J.~M. Tranquada,
  Phys. Rev. Lett. 113 (2014) 177002.

\bibitem{scal12a}
D.~J. Scalapino, 
  Rev. Mod. Phys. 84 (2012) 1383--1417.

\bibitem{esch06}
M.~Eschrig, 
  Adv. Phys. 55 (2006) 47--183.

\bibitem{lebo07}
D.~LeBoeuf, N.~Doiron-Leyraud, J.~Levallois, R.~Daou, J.-B. Bonnemaison, N.~E.
  Hussey, L.~Balicas, B.~J. Ramshaw, R.~Liang, D.~A. Bonn, W.~N. Hardy,
  S.~Adachi, C.~Proust, L.~Taillefer, 
  Nature 450 (2007) 533--536.

\bibitem{doir07}
N.~Doiron-Leyraud, C.~Proust, D.~LeBoeuf, J.~Levallois, J.-B. Bonnemaison,
  R.~Liang, D.~A. Bonn, W.~N. Hardy, L.~Taillefer, 
  Nature 447 (2007) 565--568.

\bibitem{seba08}
S.~E. Sebastian, N.~Harrison, E.~Palm, T.~P. Murphy, C.~H. Mielke, R.~Liang,
  D.~A. Bonn, W.~N. Hardy, G.~G. Lonzarich, 
  Nature 454 (2008) 200--203.

\bibitem{rigg11}
S.~C. Riggs, O.~Vafek, J.~B. Kemper, J.~B. Betts, A.~Migliori, F.~F. Balakirev,
  W.~N. Hardy, R.~Liang, D.~A. Bonn, G.~S. Boebinger, 
  Nat. Phys. 7 (2011) 332--335.

\bibitem{mill07}
A.~J. Millis, M.~R. Norman, 
  Phys. Rev. B 76 (2007) 220503(R).

\bibitem{yao11}
H.~Yao, D.-H. Lee, S.~Kivelson, 
  Phys. Rev. B 84 (2011) 012507.

\bibitem{chan10}
J.~Chang, R.~Daou, C.~Proust, D.~LeBoeuf, N.~Doiron-Leyraud, F.~Lalibert\'e,
  B.~Pingault, B.~J. Ramshaw, R.~Liang, D.~A. Bonn, W.~N. Hardy, H.~Takagi,
  A.~B. Antunes, I.~Sheikin, K.~Behnia, L.~Taillefer, 
  Phys. Rev. Lett. 104 (2010) 057005.

\bibitem{wu13a}
T.~Wu, H.~Mayaffre, S.~Kr\"amer, M.~Horvati\'c, C.~Berthier, P.~L. Kuhns, A.~P.
  Reyes, R.~Liang, W.~N. Hardy, D.~A. Bonn, M.-H. Julien, 
  Nat. Commun. 4 (2013) 2113.

\bibitem{lebo13}
D.~LeBoeuf, S.~Kramer, W.~N. Hardy, R.~Liang, D.~A. Bonn, C.~Proust,
  Nat. Phys. 9 (2013) 79--83.

\bibitem{yu14}
F.~Yu, M.~Hirschberger, T.~Loew, G.~Li, B.~J. Lawson, T.~Asaba, J.~B. Kemper,
  T.~Liang, J.~Porras, G.~S. Boebinger, J.~Singleton, B.~Keimer, L.~Li, N.~P.
  Ong, {Diamagnetic response in under-doped YBa$_2$Cu$_3$O$_{6.6}$ in high
  magnetic fields}, arXiv:1402.7371 (2014).

\bibitem{achk12}
A.~J. Achkar, R.~Sutarto, X.~Mao, F.~He, A.~Frano, S.~Blanco-Canosa,
  M.~Le~Tacon, G.~Ghiringhelli, L.~Braicovich, M.~Minola, M.~Moretti~Sala,
  C.~Mazzoli, R.~Liang, D.~A. Bonn, W.~N. Hardy, B.~Keimer, G.~A. Sawatzky,
  D.~G. Hawthorn, 
  Phys. Rev. Lett. 109 (2012) 167001.

\bibitem{blan13}
S.~Blanco-Canosa, A.~Frano, T.~Loew, Y.~Lu, J.~Porras, G.~Ghiringhelli,
  M.~Minola, C.~Mazzoli, L.~Braicovich, E.~Schierle, E.~Weschke, M.~Le~Tacon,
  B.~Keimer, 
  Phys. Rev. Lett. 110 (2013) 187001.

\bibitem{blac13a}
E.~Blackburn, J.~Chang, M.~H\"ucker, A.~T. Holmes, N.~B. Christensen, R.~Liang,
  D.~A. Bonn, W.~N. Hardy, U.~R\"utt, O.~Gutowski, M.~v. Zimmermann, E.~M.
  Forgan, S.~M. Hayden, 
  Phys. Rev. Lett. 110 (2013) 137004.

\bibitem{huck14}
M.~H\"ucker, N.~B. Christensen, A.~T. Holmes, E.~Blackburn, E.~M. Forgan,
  R.~Liang, D.~A. Bonn, W.~N. Hardy, O.~Gutowski, M.~v. Zimmermann, S.~M.
  Hayden, J.~Chang, 
  Phys. Rev. B 90 (2014) 054514.

\bibitem{blan14}
S.~Blanco-Canosa, A.~Frano, E.~Schierle, J.~Porras, T.~Loew, M.~Minola,
  M.~Bluschke, E.~Weschke, B.~Keimer, M.~Le~Tacon, 
  Phys. Rev. B 90 (2014) 054513.

\bibitem{rams12}
B.~J. Ramshaw, J.~Day, B.~Vignolle, D.~LeBoeuf, P.~Dosanjh, C.~Proust,
  L.~Taillefer, R.~Liang, W.~N. Hardy, D.~A. Bonn, 
  Phys. Rev. B 86 (2012) 174501.

\bibitem{gris14}
G.~Grissonnanche, O.~Cyr-Choini\`ere, F.~Lalibert\'e, S.~Ren\'e~de Cotret,
  A.~Juneau-Fecteau, S.~Dufour-Beaus\'ejour, M.~{\`E}. Delage, D.~LeBoeuf,
  J.~Chang, B.~J. Ramshaw, D.~A. Bonn, W.~N. Hardy, R.~Liang, S.~Adachi, N.~E.
  Hussey, B.~Vignolle, C.~Proust, M.~Sutherland, S.~Kr\"amer, J.~H. Park,
  D.~Graf, N.~Doiron-Leyraud, L.~Taillefer, {Direct measurement of the upper
  critical field in cuprate superconductors}, 
  Nat. Commun. 5 (2014) 3280.

\bibitem{huck13}
M.~H\"ucker, M.~v.~Zimmermann, Z.~J. Xu, J.~S. Wen, G.~D. Gu, J.~M. Tranquada,
  Phys. Rev. B 87 (2013) 014501.

\bibitem{comi14}
R.~Comin, A.~Frano, M.~M. Yee, Y.~Yoshida, H.~Eisaki, E.~Schierle, E.~Weschke,
  R.~Sutarto, F.~He, A.~Soumyanarayanan, Y.~He, M.~Le~Tacon, I.~S. Elfimov,
  J.~E. Hoffman, G.~A. Sawatzky, B.~Keimer, A.~Damascelli, 
  Science 343 (2014) 390--392.

\bibitem{dasi14}
E.~H. {da Silva Neto}, P.~Aynajian, A.~Frano, R.~Comin, E.~Schierle,
  E.~Weschke, A.~Gyenis, J.~Wen, J.~Schneeloch, Z.~Xu, S.~Ono, G.~Gu,
  M.~Le~Tacon, A.~Yazdani, 
  Science 343 (2014) 393--396.

\bibitem{tabi14}
W.~Tabis, Y.~Li, M.~L. Tacon, L.~Braicovich, A.~Kreyssig, M.~Minola, G.~Dellea,
  E.~Weschke, M.~J. Veit, M.~Ramazanoglu, A.~I. Goldman, T.~Schmitt,
  G.~Ghiringhelli, N.~Bari{\v s}i\'c, M.~K. Chan, C.~J. Dorow, G.~Yu, X.~Zhao,
  B.~Keimer, M.~Greven, {Connection between charge-density-wave order and
  charge transport in the cuprate superconductors}, arXiv:1404.7658 (2014).

\bibitem{dasi14b}
E.~H. {da Silva Neto}, R.~Comin, F.~He, R.~Sutarto, Y.~Jiang, R.~L. Greene,
  G.~A. Sawatzky, A.~Damascelli, {Charge ordering in the electron-doped
  superconductor Nd$_{2-x}$Ce$_x$CuO$_4$}, arXiv:1410.2253 (2014).

\bibitem{howa03b}
C.~Howald, H.~Eisaki, N.~Kaneko, M.~Greven, A.~Kapitulnik, Phys. Rev. B 67
  (2003) 014533.

\bibitem{kohs07}
Y.~Kohsaka, C.~Taylor, K.~Fujita, A.~Schmidt, C.~Lupien, T.~Hanaguri, M.~Azuma,
  M.~Takano, H.~Eisaki, H.~Takagi, S.~Uchida, J.~C. Davis, 
  Science 315~(5817) (2007) 1380--1385.

\bibitem{park10}
C.~V. Parker, P.~Aynajian, E.~H. da~Silva~Neto, A.~Pushp, S.~Ono, J.~Wen,
  Z.~Xu, G.~Gu, A.~Yazdani, 
  Nature 468 (2010) 677--680.

\bibitem{enok13}
M.~Enoki, M.~Fujita, T.~Nishizaki, S.~Iikubo, D.~K. Singh, S.~Chang, J.~M.
  Tranquada, K.~Yamada, 
  Phys. Rev. Lett. 110 (2013) 017004.

\bibitem{wu14}
T.~Wu, H.~Mayaffre, S.~Kr\"amer, M.~Horvati\'c, C.~Berthier, W.~N. Hardy,
  R.~Liang, D.~A. Bonn, M.-H. Julien, {Short-range charge order reveals the
  role of disorder in the pseudogap state of high-$T_c$ superconductors},
  arXiv:1404.1617 (2014).

\bibitem{bobr97}
J.~Bobroff, H.~Alloul, P.~Mendels, V.~Viallet, J.-F. Marucco, D.~Colson,
  Phys. Rev. Lett. 78 (1997) 3757--3760.

\bibitem{hoff02}
J.~E. Hoffman, E.~W. Hudson, K.~M. Lang, V.~Madhavan, H.~Eisaki, S.~Uchida,
  J.~C. Davis, 
  Science 295 (2002) 466--469.

\bibitem{hink08}
V.~Hinkov, D.~Haug, B.~Fauqu\'e, P.~Bourges, Y.~Sidis, A.~Ivanov, C.~Bernhard,
  C.~T. Lin, B.~Keimer, 
  Science 319 (2008) 597.

\bibitem{haug09}
D.~Haug, V.~Hinkov, A.~Suchaneck, D.~S. Inosov, N.~B. Christensen,
  C.~Niedermayer, P.~Bourges, Y.~Sidis, J.~T. Park, A.~Ivanov, C.~T. Lin,
  J.~Mesot, B.~Keimer, 
  Phys. Rev. Lett. 103 (2009) 017001.

\bibitem{frad12}
E.~Fradkin, S.~A. Kivelson, 
   Nat. Phys. 8 (2012) 864--866.
   
\bibitem{berg09a}
E.~Berg, E.~Fradkin, and S.~A.~Kivelson,
Nat. Phys. 5 (2009) 830--833.

\end{thebibliography}


\end{document}